\documentclass[twocolumn,showpacs,preprintnumbers,amsmath,amssymb,superscriptaddress]{revtex4}

\usepackage[dvips]{color}
\usepackage{graphicx}% Include figure files
\usepackage{dcolumn}% Align table columns on decimal point
\usepackage{bm}% bold math
\usepackage{dsfont}% nice one operators

\definecolor{lightblue}{rgb}{0.7, 0.7, 1}
\definecolor{lightred}{rgb}{1, 0.8, 0.8}
\definecolor{red}{rgb}{1, 0, 0}
\definecolor{blue}{rgb}{0, 0, 1}

\begin{document}
\bibliographystyle{unsrt}

\title{Oscillations and interactions of dark and dark-bright solitons in Bose-Einstein condensates}

\author{C. Becker}
\affiliation{Institut f{\"u}r Laserphysik, Universit{\"a}t
Hamburg, Luruper Chaussee 149, D-22761 Hamburg, Germany.}
\author{S. Stellmer}
\affiliation{Institut f{\"u}r Laserphysik, Universit{\"a}t
Hamburg, Luruper Chaussee 149, D-22761 Hamburg, Germany.}
\author{P. Soltan-Panahi}
\affiliation{Institut f{\"u}r Laserphysik, Universit{\"a}t
Hamburg, Luruper Chaussee 149, D-22761 Hamburg, Germany.}
\author{S. D\"orscher}
\affiliation{Institut f{\"u}r Laserphysik, Universit{\"a}t
Hamburg, Luruper Chaussee 149, D-22761 Hamburg, Germany.}
\author{M. Baumert}
\affiliation{Institut f{\"u}r Laserphysik, Universit{\"a}t
Hamburg, Luruper Chaussee 149, D-22761 Hamburg, Germany.}
\author{E.-M. Richter }
\affiliation{Institut f{\"u}r Laserphysik, Universit{\"a}t
Hamburg, Luruper Chaussee 149, D-22761 Hamburg, Germany.}
\author{J. Kronj{\"a}ger}
\affiliation{Institut f{\"u}r Laserphysik, Universit{\"a}t
Hamburg, Luruper Chaussee 149, D-22761 Hamburg, Germany.}
\author{K. Bongs}
\affiliation{MUARC, School of Physics and Astronomy, University of Birmingham, Edgebaston, Birmingham B15 2TT, UK}
\author{K. Sengstock}
\affiliation{Institut f{\"u}r Laserphysik, Universit{\"a}t Hamburg,
  Luruper Chaussee 149, D-22761 Hamburg, Germany.}

\date{\today}% It is always \today, today, but any date may be explicitly specified

\begin{abstract}
Solitons are among the most distinguishing fundamental excitations in a wide range of non-linear systems such as
water in narrow channels, high speed optical communication, molecular biology and
astrophysics. Stabilized by a balance between spreading and focusing, solitons are
wavepackets, which share some exceptional generic features like form-stability and
particle-like properties. Ultra-cold quantum gases represent very pure and well-controlled non-linear systems,
therefore offering unique possibilities to study soliton dynamics.
Here we report on the first observation of long-lived dark and dark-bright solitons with lifetimes of up to
several seconds as well as their dynamics in highly stable optically trapped $^{87}$Rb
Bose-Einstein condensates. 
In particular, our detailed studies of dark and dark-bright soliton oscillations reveal the particle-like 
nature of these collective excitations for the first time. 
In addition, we discuss the collision between these two types of solitary excitations in Bose-Einstein condensates.
\end{abstract}

% PACS, the Physics and Astronomy Classification Scheme.
%\keywords{Suggested keywords}%Use showkeys class option if keyword display desired
\maketitle

The dynamics of non-linear systems plays an essential role in nature,
ranging from strong non-linear interactions of elementary particles 
to non-linear wave phenomena in oceanography and meteorology.
A special class of non-linear phenomena are solitons with interesting particle
and wave-like behaviour. Reaching back to the observation of "waves of translation" in a
narrow water channel by Scott-Russell in 1834 \cite{ScottRussell1849}, solitons are nowadays
recognized to appear in various systems as different as astrophysics, molecular biology and 
non-linear optics \cite{Kivshar1998}. 
They are characterized as localized solitary wavepackets that maintain their 
shape and amplitude caused by a self-stabilization against dispersion via 
a non-linear interaction.
While an early theoretical explanation of this non-dispersive wave phenomenon was given 
by Korteweg and de Vries in the late 19th century it was not before 1965 that numerical
simulations of Zabusky and Kruskal theoretically proved that these solitary waves preserve their 
identity in collisions \cite{Zabusky1965,Zabusky1968}.
This revelation led to the term "soliton" for this type of collective excitation.\\
Nowadays solitons are a very active field of research in many areas of science.
In the field of non-linear optics they attract enormous attention
due to applications in fast data transfer.\\
Bose-Einstein condensates (BEC) of weakly interacting atoms offer fascinating
possibilities for the study of non-linear phenomena, as they are very pure samples of ultra-cold
gases building up an effective macroscopic wave function of up to mm size.
Non-linear effects like collective excitations \cite{Mewes1996}, four-wave mixing \cite{Deng1999}
and vortices \cite{Matthews1999,Madison2000} have been studied, to name only a few examples.
The existence and some fundamental properties of solitons have been deduced from 
few experiments employing ultra-cold quantum gases.
Bright solitons, characterized as non-spreading matter-wave packets, have been 
observed in BEC with attractive interaction
\cite{Strecker2002,Khaykovich2002,Cornish2006} where they represent the ground state 
of the system. 
In a repulsively interacting condensate confined in a periodic potential, bright gap solitons 
have been realized \cite{Eiermann2003,Eiermann2004} by modelling a suitable anomalous dispersion.\\
BEC with repulsive interactions allow for dark soliton solutions, characterized by 
a notch in the density distribution. 
In contrast to bright solitons, dark solitons are truly excited states with energies 
greater than the underlying BEC ground state. 
Dark solitons have been generated in few pioneering experiments 
\cite{Burger1999,Denschlag2000,Anderson2001} boosting an immense theoretical interest 
in these non-linear structures in BEC. 
Dynamical \cite{Reinhardt1997,Jackson1998,Anderson2001,Feder2000,Brand2002,Muryshev1999,Busch2000,Theocharis2007} 
and thermodynamical \cite{Jackson2007,Fedichev1999,Muryshev2002} instabilities as well as 
collisional properties \cite{Carr2001,Burger2002} have been analyzed theoretically  
in great detail. 
Moreover, the existence of bright solitons stabilized by the presence of dark solitons 
in another quantum state in repulsive BEC has been proposed \cite{Busch2000} 
and confirmed in a proof-of-principle experiment \cite{Anderson2001}.
The occurence of undamped oscillations of solitons in axially harmonic traps comparable 
to a single particle with a large negative mass has been predicted by mean field theory 
\cite{Busch2000,Fedichev1999} and supported by numerical simulations \cite{Jackson2007} 
for dark solitons and appears as one of the paradigms of soliton physics
in BEC as it clearly demonstrates the particle character of the soliton.
Limited by very short lifetimes in previous experiments however, oscillations of dark 
solitons have not been observed yet.\\
In this paper, we show that very long-lived dark solitons can be generated in BEC, facilitating 
detailed studies of soliton oscillations as well as the creation of dark-bright solitons by 
filling the dark soliton with atoms in another hyperfine state. 
In addition the reflection of a dark soliton off a filled soliton has been observed and is presented. 
In the following we give a brief introduction to the description of solitons in BEC and
characteristic properties of soliton dynamics.
Close to absolute zero temperature, where thermal fluctuations can be neglected to first order, the 
condensate is well described within the framework of the non-linear Gross-Pitaevskii equation (GPE) 
\cite{Pitaevskii2003}, which is known to support soliton solutions as it is closely related 
to the cubic non-linear equation vastly employed in non-linear optics to describe 
solitary wave propagation in optical fibers \cite{Kivshar1998}
\begin{eqnarray}
i \hbar \dot{\psi}(z,t) =	&-&\frac{\hbar^2}{2m}\psi''(z,t)\nonumber\\
				&+& \left[ V_{ext}(z) +g \left|\psi(z,t)\right|^2\right]\psi(z,t).
\end{eqnarray}
Here $\psi$ denotes the condensate wavefunction, $g=2\hbar \omega_{\bot}a$ is a measure of the non-linear 
atomic interaction  in quasi-1D geometries with strong transverse confinement. $g$ is determined by the 
s-wave scattering length $a$ in quasi-1D and the transverse trapping frequency $\omega_{\bot}$. 
$V_{ext}$ describes a confining external potential.
A dark soliton solution to the GPE of a homogeneous BEC describing a density notch at position 
$q$ propagating along $z$ with a velocity $\dot{q}$ can be written as \cite{Shabat1972}
\begin{eqnarray}
\label{equ:ds}
\psi_{D}(z,t)&=&\sqrt{n_{0}}\Bigg\{i \frac{\dot{q}}{\bar{c}_{s}} + \sqrt{1-\frac{\dot{q}^{2}}{\bar{c}_{s}^{2}}}
\,\tanh \left[ \kappa\,(z-q(t)) \right]\Bigg\}\nonumber\\
& &\times e^{-ign_{0}t/\hbar},
\end{eqnarray}
where the speed of sound in a quasi-1D condensate is given by $\bar{c}_{s}=\sqrt{n_{0}g/2m}$. 
$n_{0}$ is the peak density of the condensate. The inverse size of the soliton $\kappa$ 
is determined by the healing length $\bar{\xi} = \hbar/m\bar{c}_{s}$ and the soliton speed $\dot{q}$ through 
$\kappa=\bar{\xi}^{-1}\times \sqrt{1-(\dot{q}/\bar{c}_{s})^{2}}$. 
Note that in a quasi-1D condensate, the averaging of the density over the radial degrees 
of freedom effectively changes the speed of sound to $\bar{c}_{s}=c_{s}/\sqrt{2}$ and 
the healing length to $\bar{\xi}=\sqrt{2}\,\xi$ as compared to their 3D values $c_{s}$ and $\xi$ respectively.
The phase and density distributions of a dark soliton are schematically shown in Fig.\,1b.
As depicted, the phase only shows significant changes in the vicinity of the nodal plane of 
the soliton and is constant otherwise. 
Crossing the nodal plane of the soliton, the wavefunction accumulates a specific phase slip 
between $0$ and $\pi$ depending on the depth and speed of the soliton related by 
$n_{s}/n_{0}=1-(\dot{q}/\bar{c}_{s})^{2} = \sin^2(\phi/2)$ where $n_{s}$ denotes the missing
density at the position of the soliton. 
A phase jump of $\Delta \phi=\pi$ corresponds to a fully modulated soliton with zero 
velocity representing the only time independent soliton solution of the GPE. 
As the phase difference diminishes the soliton velocity grows while it gets shallower 
and wider ultimately vanishing at the speed of sound.
The preceeding considerations have led to the idea and experimental realization of 
creating dark solitons in BEC by optically imprinting a phase gradient of approximately 
$\pi$ over a spatial region not larger than the healing length 
\cite{Dobrek1999,Burger1999,Denschlag2000}.\\
While solitons in quasi one dimensional BEC are dynamically stable \cite{Jackson1998,Fedichev1999}, 
experiments \cite{Denschlag2000,Anderson2001} and intensive theoretical studies suggest that 
in less constrained geometries, where the condition $\gamma=n_{0}g/\hbar \omega_{\bot} \ll 1 $ 
is not perfectly met, the growth of dynamically unstable modes will lead to a transfer of kinetic energy 
of the soliton to radial excitation modes of the condensate, mediated by the atomic interaction 
and resulting in a bending of the soliton plane \cite{Denschlag2000}, which may ultimately decay into 
vortex pairs as reported in Ref. \cite{Anderson2001}. 
Since the energy of a dark soliton is always greater than the energy of the ground state condensate 
it is thermodynamically unstable in any case and shows fast decay even at reasonable low temperatures 
\cite{Muryshev2002,Jackson2007,Parker2003} as observed in experiment \cite{Burger1999,Denschlag2000}. 
The dissipation accelerates the soliton according to its negative kinetic energy 
until it vanishes and smoothly transforms to the BEC ground state as it approaches the speed 
of sound \cite{Fedichev1999}. 
This interesting aspect can be interpreted as an accelerating instability \cite{Busch2000} 
and implies that a negative mass can be assigned to a dark soliton.
Lifetimes on the order of $10\,\mathrm{ms}$ have been reported preventing the observation of more complex soliton 
physics like oscillations or collisions.\\
We have developed a reliable, robust method to produce elongated $^{87}$Rb-BEC at extremely low
temperatures in an optical trapping potential overcoming former technical limitations.
We produce a BEC composed of $5 \times 10^4 \,\,^{87}$Rb atoms in the $5^2S_{1/2},\, F=1,\, m_F=-1$ state 
in an optical dipole trap with trapping frequencies $\omega_{z}=2\pi \times 5.9\,\mathrm{Hz}$,
$\omega_{\bot}^{\mathrm{ver}}=2\pi \times 85\,\mathrm{Hz}$ 
and $\omega_{\bot}^{\mathrm{hor}}=2\pi \times 133\,\mathrm{Hz}$ with no discernable thermal fraction.
Trap frequencies have been cross-checked by the measurement of various collective oscillations.
Typical atomic peak densities are $5.8 \times 10^{13}$ cm$^{-3}$ implying a speed of sound of 
$\bar{c}_{s}=1.0\,$mm/s. 
The chemical potential is on the order of $20\,\mathrm{nK}$.\\
Ultra-stable laboratory conditions ensure an exceptional reproducibility enabling us to record 
time series of soliton dynamics with unprecedented precision. 
The low trap depth guarantees a slight but constant evaporative cooling, so that no heating can be detected 
for time scales as large as the lifetime of the condensate, which is greater than $10\,\mathrm{s}$.
Solitons are produced by optically imprinting a phase gradient as shown in Fig.\,1a: a part of the 
condensate is exposed to the dipole potential $U_{\mathrm{dip}}$ of a laser beam detuned by some tens of GHz from atomic resonance. 
We image an optical mask pattern onto the BEC with diffraction-limited optical resolution of better than 
$2\,\mu\mathrm{m}$. This results in a phase evolution of the masked relative to the unmasked part of the condensate 
of $\Delta \phi = U_{\mathrm{dip}}t/(i\hbar) $.
The pattern is generated by a spatial light modulator (SLM) with an effective pixel size of 
$0.8 \,\mu\mathrm{m}$ allowing for almost arbitrary optical potentials \cite{Boyer2006}. 
To imprint a phase slip of order $\pi$, we choose a pulse time $t_{\pi}=40\,\mu\mathrm{s}$, much smaller than 
the correlation time $\tau_{\mathrm{corr}}=\bar{\xi}/\bar{c}_{s}=700\,\mu$s for our experimental parameters 
to avoid a simultaneous disturbance of the atomic density.
This phase gradient leads to a local superfluid velocity of the condensate according to 
$v_{SF}=\hbar/m \partial_{z} \phi$ which can also be interpreted as a local potential gradient 
transferring momentum to the BEC, thus assisting the formation of a density minimum \cite{Burger1999}.
Since a dark soliton can be regarded as a hole rather than a particle it moves in the direction 
opposite to the superfluid flow of the condensate. 
The appropriate equation of motion for small velocities 
$\dot{q}$ reads $M_{s}\,\ddot{q}(t)=-1/2 \,\partial_{z}V(z)$ where $M_s$ is the negative mass of the soliton.
This implies a soliton oscillation frequency of $\omega = \omega_{z}/\sqrt{2}$ for harmonic 
confinement \cite{Fedichev1999,Busch2000}.\\
Fig.\,2a shows the time evolution of a dark soliton created by the aforementioned phase 
imprinting method. Absorption images were taken after a time-of-flight of $11.5\,\mathrm{ms}$ to allow 
the condensate and soliton to expand since the soliton size $l_{s}\approx\bar{\xi}\approx0.8\,\mu\mathrm{m}$ 
in the trap is beyond optical resolution. 
The soliton clearly propagates axially along the condensate with an initial veloctiy of 
$\dot{q}=0.56\,\mathrm{mm/s}=0.56 \,\bar{c}_{s}$ indicating a relative soliton depth of $0.68$. 
We were able to detect nearly pure dark solitons after times as long as $2.8\,\mathrm{s}$ in single
experimental realizations (see Fig.\,1e), surpassing lifetimes of dark solitons in any former experimental
realization by more than a factor of 200.
Fluctuations in the soliton position due to small preparation errors however 
prevent the observation of soliton oscillations for evolution times $\tau_{\mathrm{evol}} \gg 250\,\mathrm{ms}$.   
The extraordinary long lifetimes facilitate the first observation of an oscillation of a dark soliton in a trapped BEC. 
An oscillation period of $\Omega=2\pi \times (3.8\pm 0.1)\,\mathrm{Hz}$ has been recorded and could be followed for more than one period. 
In contrast to the theoretical prediction of $\omega_{z}/\sqrt2 = 2\pi \times 4.2\,\mathrm{Hz}$ for a dark soliton in an axially 
harmonic trap the observed period indicates that the harmonic approximation for the trapping potential starts to break down. 
Caused by the shallowness of our dipole trap the atoms rather experience a Gaussian potential, which is less steep 
than harmonic leading to a larger amplitude-dependent oscillation period for the soliton. 
We have calculated the oscillation frequency in a Gaussian potential created by a laser beam with a waist of 
$125\,\mu\mathrm{m}$ with the observed soliton amplitude of $Z_{s}=33\,\mu\mathrm{m}$ and find an oscillation
frequency of $4.0\,\mathrm{Hz}$. This is in good agreement with our experimental data. 
Furthermore, the observed amplitude allows for a consistency check of the soliton depth. 
At the turning point of the soliton motion $Z_{s}$ the constant soliton depth equals the 
Thomas-Fermi density $n_{TF}(Z_{s})$ of the condensate and interrupts the superfluid flow of atoms.
At this point the soliton starts to move in the opposite direction. 
Given the measured initial speed of the soliton and the observed density distribution of the condensate 
$Z_{s}$ can be calculated to be $36\,\mu\mathrm{m}$ and is in very good agreement with the measured value.\\
Another feature extracted from Fig.\,2a is a density wave that travels in the opposite 
direction at a velocity equal to the speed of sound.
The occurrance of such a density wave has been investigated theoretically and experimentally \cite{Burger1999}
and has been attributed to the method of soliton generation via phase imprinting while 
leaving the instantaneous density distribution unchanged.
The density waves die out after approximately $50\,\mathrm{ms}$ leaving a flat BEC with only
one soliton excitation. \\
Calculating the dimensionality parameter $\gamma:=n_{0}g/\hbar \omega_{\bot} = 3.7$ and comparing 
this value to the critical ratio $\gamma_{c}$ given by  Muryshev {\it et al.} \cite{Muryshev2002} 
we find our soliton to be right on the edge of the region of dynamical stability.
This is confirmed regarding the observed soliton lifetimes. \\ 
We have performed numerical simulations of the GPE showing that the phase imprinting method cannot create
single perfect dark solitons but always creates density waves that carry away part of the imprinted 
phase gradient. 
Moreover the occurence of a second small soliton can be extracted from the simulations as shown
in Fig.\,2b.\\
The crucial feature to the observed long lifetimes of dark solitons seems to be the very low temperature of our samples. 
The critical temperature for Bose-Einstein condensation for our experimental parameters 
is $(67 \pm 5)\,\mathrm{nK}$.
Estimating that a thermal fraction of at least 10\% could have been detected in absorbtion imaging 
--which was not the case-- an upper limit for the temperature of $T \leq 0.5 \,T_{c}=30\,\mathrm{nK}$ 
can be given which is on the order of the chemical potential $\mu$.
We assume a significantly smaller temperature, since temperatures of $T \approx 0.2 \, T_{c}$ 
would already considerably limit the solitons lifetime \cite{Jackson2007}, which has not been observed in our 
experiment.\\
Despite the interesting physics that can be investigated using dark solitons, so called dark-bright solitons 
appearing in multi-component BEC show even more fascinating physical properties, such as enhanced dynamical 
stability and the possibility of bright-component particle exchange in soliton collisions.
A dark-bright soliton is basically a dark soliton filled with atoms of a different species or in another
internal state of an atomic matter wave. 
A dark-bright soliton can be generated in a $^{87}$Rb BEC by imprinting a dark soliton in state
$\left|F=1,m_{F}=0\right\rangle$ and filling the density dip with atoms in state $\left|F=2,m_{F}=0\right\rangle$ 
\cite{Dum1998,Busch2001} leading to the density distribution depicted in Fig.\,3b. 
While dark solitons are unstable to transverse excitations with wavelengths greater than their 
extension $l_{s}\approx\bar{\xi}$ large dark-bright solitons are expected to overcome this 
restriction since their size can be much larger than $\bar{\xi}$ when the number of atoms in the other hyperfine
state becomes very large. Dark-bright solitons should therefore be robust in traps geometries, 
which are not truly 1D.
We have employed a method to simultaneously imprint the phase gradient and transfer atoms to 
the other hyperfine state via a coherent Raman pulse technique using a laser system phase-locked 
on the two-photon hyperfine resonance $\left|F=1,m_{F}=0\right\rangle \rightarrow \left|F=2,m_{F}=0\right\rangle$
(see Fig.\,3a). 
Performing a $2\pi$ pulse of duration $40 \,\mu\mathrm{s}$ on one side of the condensate leaves the population 
effectively unchanged, but introduces a phase difference of $\pi$ compared to the unperturbed part of 
the condensate. In a small region around the edge of the mask $l\approx \xi$, a transfer of the 
population to the $\left|F=2,m_{F}=0\right\rangle$ state occurs (Fig.\,3b). 
By employing a step-like intensity pattern rather than a simple edge it is possible to vary the number 
of atoms transferred to the other hyperfine state by changing the width of the intermediate step.
The mask pattern used in the experiment described here resulted in a bright component population 
$N_{B} = 0.08N_{\mathrm{tot}}$, where $N_{\mathrm{tot}}$ is the total number of atoms.
A time series of the propagation of such a dark-bright soliton is shown in Fig.\,3c. 
After a short time of flight of $9\,\mathrm{ms}$ the atoms are first exposed to a light pulse resonant only with 
$\left|F=2\right\rangle$. 
After another $2\,\mathrm{ms}$ the $\left|F=1\right\rangle$ atoms are subsequently imaged.
The dynamics of the dark-bright soliton could be followed for more than 2s as seen in Fig.\,3d.
We observe an oscillation with a frequency of $\Omega_{\mathrm{db}}=2\pi \times (0.90\pm 0.02)\,\mathrm{Hz}=0.24\times \Omega$ 
much smaller than the frequency of the corresponding dark soliton. 
An expression for the oscillation frequency can be given employing the equation of motion for very strongly 
populated dark-bright solitons \cite{Busch2001} and leads to
\begin{equation}
\Omega_{\mathrm{db}}=\Omega \,\alpha_{z_{\mathrm{max}}} \, \frac{2N_{\mathrm{tot}}\,\bar{\xi}}{N_{\mathrm{b}}\,R_{z}}. 
\end{equation}
where $R_{z}$ denotes the radius of the BEC along the axial direction and $\alpha_{z_{\mathrm{max}}}$ is an
amplitude-dependent numerical factor which takes into account, that the total potential experienced by the bright component 
depends on the inhomogeneous density of the dark component.
For the observed values of $N_{\mathrm{b}}, R_{z}$ and $N_{\mathrm{tot}}$ we calculate an oscillation frequency of 
$1.27\,\mathrm{Hz}$ which is in very good agreement with the observed value.\\
Another spectacular feature that can be extracted from this measurement is the interaction of a dark soliton
with the much slower dark-bright soliton. Owing to the method of initial state preparation, an extra dark soliton
is always generated in addition to the dark-bright soliton. 
As shown in Fig.\,4 the dark soliton propagates in the opposite direction as compared to the dark-bright
one and oscillates back with the same frequency as a dark soliton in an unperturbed experiment. 
After $120\,\mathrm{ms}$ it thus approaches the position of the dark-bright soliton which 
has only moved very little due to its much smaller oscillation frequency. 
The dark soliton is reflected off the dark-bright one comparable to a hard-wall reflection and moves back.
To our knowledge this is the first observation of collisions of different types of matter wave solitons.
In conclusion we have realized long-lived solitons in $^{87}$Rb BEC and observed soliton oscillations for the first time.
Via a combination of a local Raman transfer and phase imprinting method we could realize two-component solitary excitations
called dark-bright solitons which clearly exhibit very slow oscillatory dynamics in a trapped BEC.
As a first striking example for soliton interaction the reflection of a dark soliton bouncing off a filled soliton could be
observed.
These experiments pave the way for further studies on solitons and soliton dynamics in ultracold quantum gases.

\subsection*{Methods}
\subsubsection*{Creation of solitons}
We create Bose-Einstein condensates of $^{87}$Rb atoms via trapping of up to $5\times10^{9}$ atoms in a 3D
magneto-optical trap (MOT), sub-doppler cooling of these atoms and subsequently transferring them into a 
magnetic trap.
We evaporatively cool the atoms slightly above the critical temperature $T_{c}$ for Bose-Einstein condensation 
within $20\,\mathrm{s}$.
Afterwards we superimpose a crossed dipole trap realized by a Nd:YAG laser beam focused to a waist of $35\,\mu\mathrm{m}$
and a perpendicular Ti:Sa laser beam ($830\,\mathrm{nm}$) focused to a waist of $125\,\mu\mathrm{m}$.
The atoms are loaded in this dipole trap and  further cooled evaporatively by smoothly lowering the optical power of 
the dipole trap beams until we end up with an almost pure BEC of $5-10\times 10^{4}$ atoms. 
We reduce the dipole trap power as much as possible in order to ensure smallest temperatures of the BEC.
The trapping frequencies for this configuration have been determined by several independent approaches 
and read $2\pi\times(6.3, 85, 133)\,\mathrm{rad\,s^{-1}}$.
At this stage the condensate consists of atoms in the $\left|F=1,m_{F}=-1\right\rangle$ state and can 
be transferred to any other state or superposition of states via rf- or microwave-pulse or -sweep techniques. 
The state preparation is carried out at a sufficiently large magnetic offset field to avoid undesired spin 
mixing dynamics \cite{Kronjaeger2006}. 
Moreover, the creation of dark-bright solitons demands a spatially selective transfer to the 
$\left|F=2,m_{F}=0\right\rangle$ state which is accomplished by the use of a phase-locked Raman laser system 
with a relative phase error of not more than 0.44 rad. 
The optical transfer and phase imprinting is achieved by imaging a computer-generated pattern displayed on a  
spatial light modulator with a pixel size of $8\,\mu\mathrm{m}$ onto the BEC with an optical resolution 
of better than $2\,\mu\mathrm{m}$. 
This is achieved via a high-quality imaging optics with a magnification of $1/10$. This optics is also used for the 
detection of BEC then yielding a magnification of 10.
After creation of BEC in the optical dipole trap and the application of the phase imprinting light pattern
via the SLM for a duration of $40\,\mu\mathrm{s}$, we allow the BEC to evolve in the trap for a variable time 
$\tau_{\mathrm{evol}}$.
Subsequently, we switch off the optical trapping potential within $1\,\mu\mathrm{s}$.
In the experiments presented here we take an absorption image of the expanded atomic cloud
after a time-of-flight of $11\,\mathrm{ms}$.
Note that this is a destructive detection technique. 
For the measurements presented here, each data point therefore corresponds to a new realization of BEC, 
phase imprinting, evolution time and detection.
\subsubsection*{Determination of soliton parameters}
We have determined soliton parameters such as position, width and amplitude from 2D-fits to the absorption
images. The employed function consists of a Thomas-Fermi like density distribution for the BEC modulated
with individual solitons basically given by the square of Eqn. \ref{equ:ds}.
The bright component has been fitted employing a function derived from the wavefunction of the bright component
of the dark-bright soliton \cite{Busch2001}
 \begin{equation}
\psi_B(z(t))=\sqrt{\frac{N_B\kappa}{2}}\, \textrm{sech}\{\kappa(x-q(t))\}.
\end{equation}

\subsection*{Acknowledgements}

We thank the Deutsche Forschungsgemeinschaft DFG for funding within the Forschergruppe FOR801.
K. B. thanks EPSRC for financial support in grant EP/E036473/1.

\subsection*{Competing financial interest}

The authors declare no competing financial interests.

\clearpage

\newpage

\begin{figure}[f]
  \centering
  \includegraphics[width=\textwidth]{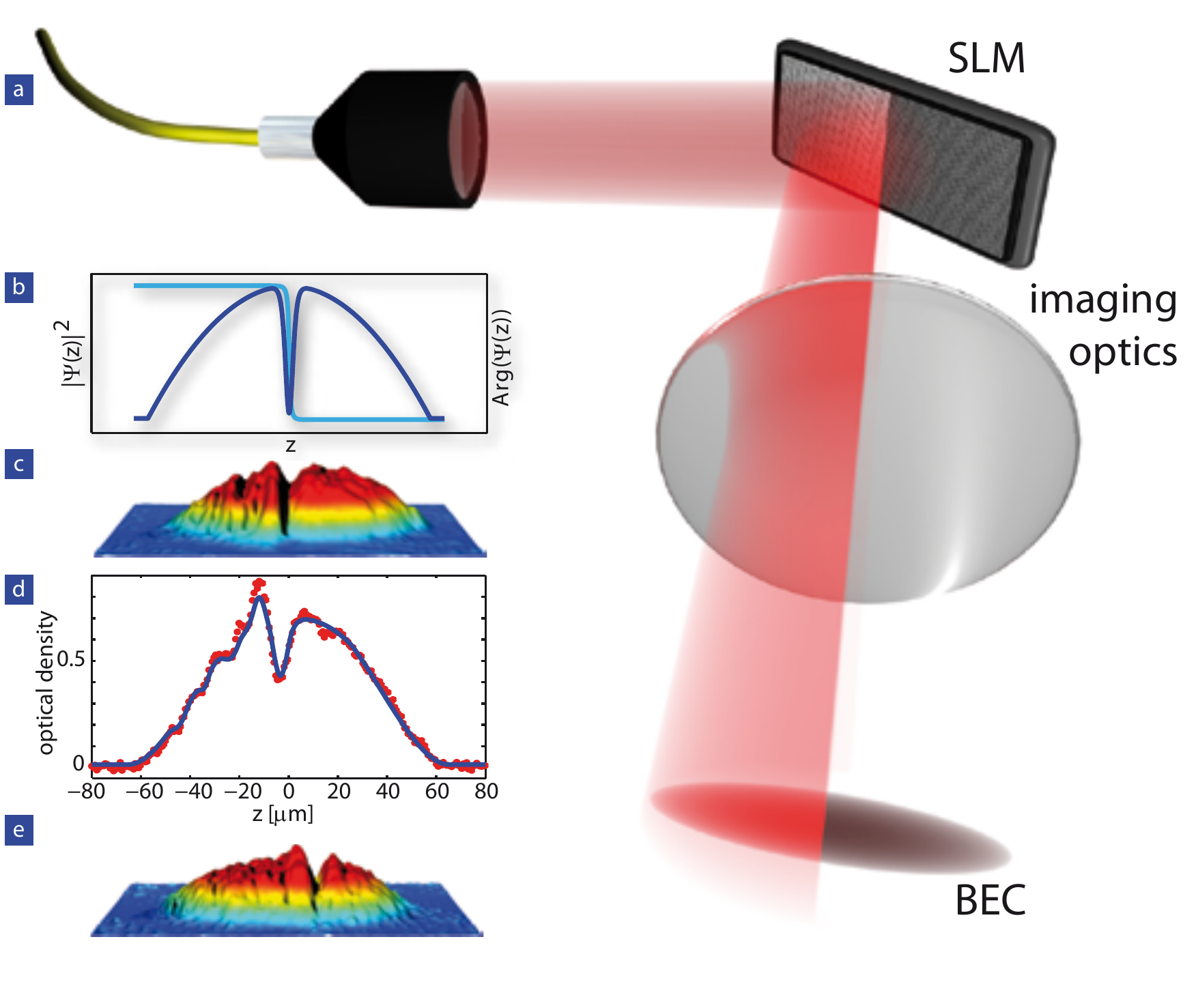}\\
  \caption[Figure1]{
	\textbf{Figure 1 | Principle of dark soliton generation.}
	\textbf{a} Optical setup: A spatial light modulator (SLM) is used to imprint a phase step
	by exposing part of the condensate to a far-detuned laser beam.
	\textbf{b} Theoretical curve of a dark soliton's density $|\Psi|^2$ ({\color{blue}\rule[2pt]{10pt}{2pt}}) 
	and phase $\phi$ ({\color{lightblue}\rule[2pt]{10pt}{2pt}}), as described by Eqn. 2. 
	\textbf{c} A typical absorption image of the condensate, taken directly after preparation of the soliton 
	and a subsequent free expansion of  $11\,\mathrm{ms}$. 
	Optical density is color- and height-coded for better visibility.
	\textbf{d} Integrated column density ({\color{red}$\bullet$}) of the data in c together with a fit 
	to the data ({\color{blue}\rule[2pt]{10pt}{2pt}}).
	\textbf{e} Image of a soliton after an evolution time of $2.8\,\mathrm{s}$.}
   \label{Figure1}
\end{figure}

\clearpage

\newpage

\begin{figure}[f]
  \centering
  \includegraphics{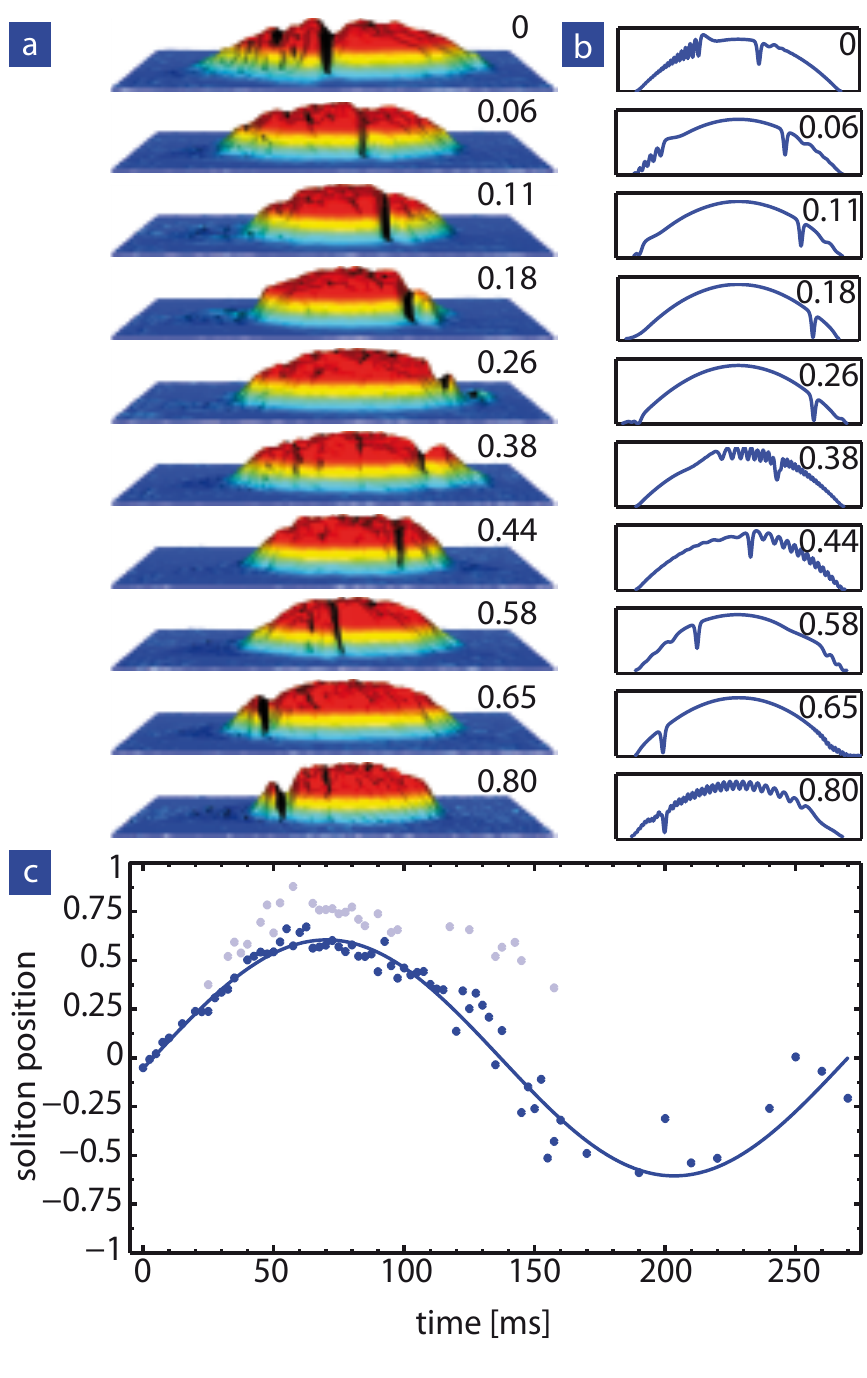}\\
  \caption[Figure2]{
	\textbf{Figure 2 | Dark soliton oscillations in a trapped BEC.}
	\textbf{a} A set of absorption images showing the soliton position at various times after phase imprinting. 
	The soliton propagates to the right and is reflected off the edge of the condensate after $t\approx$ 80 ms. 
	The corresponding evolution time for each image is given in units of the oscillation period $T$.
	\textbf{b} Results of a numerical calculation solving the 1D Gross-Pitaevski equation corresponding to our 
	parameters in units of $T$ are presented. 
	Experimentally observed features like density modulations caused by a density wave on the left side of the condensate
	as well as the development of a tiny second soliton are reproduced. 
	\textbf{c} Axial positions of the soliton ({\color{blue}$\bullet$}) with respect to the center of mass  
	and normalized to the width of the condensate. 
	The oscillation frequency is $\Omega=2\pi \times (3.8\pm 0.1)\,\mathrm{Hz}$. 
	The positition of a second tiny soliton ({\color{lightblue}$\bullet$}) as well as a sinusoidal fit 
	({\color{blue}\rule[2pt]{10pt}{2pt}}) to the position of the soliton are shown. 
	Each data point was obtained from a different experimental run. 
	The scatter is due to small fluctuations in the preparation process.
	Errors in extracting the solitons position from the individual images are typically less than 0.02 and 
	therefore not plotted.}
   \label{Figure2}
\end{figure}

\begin{figure}[f]
  \centering
  \includegraphics{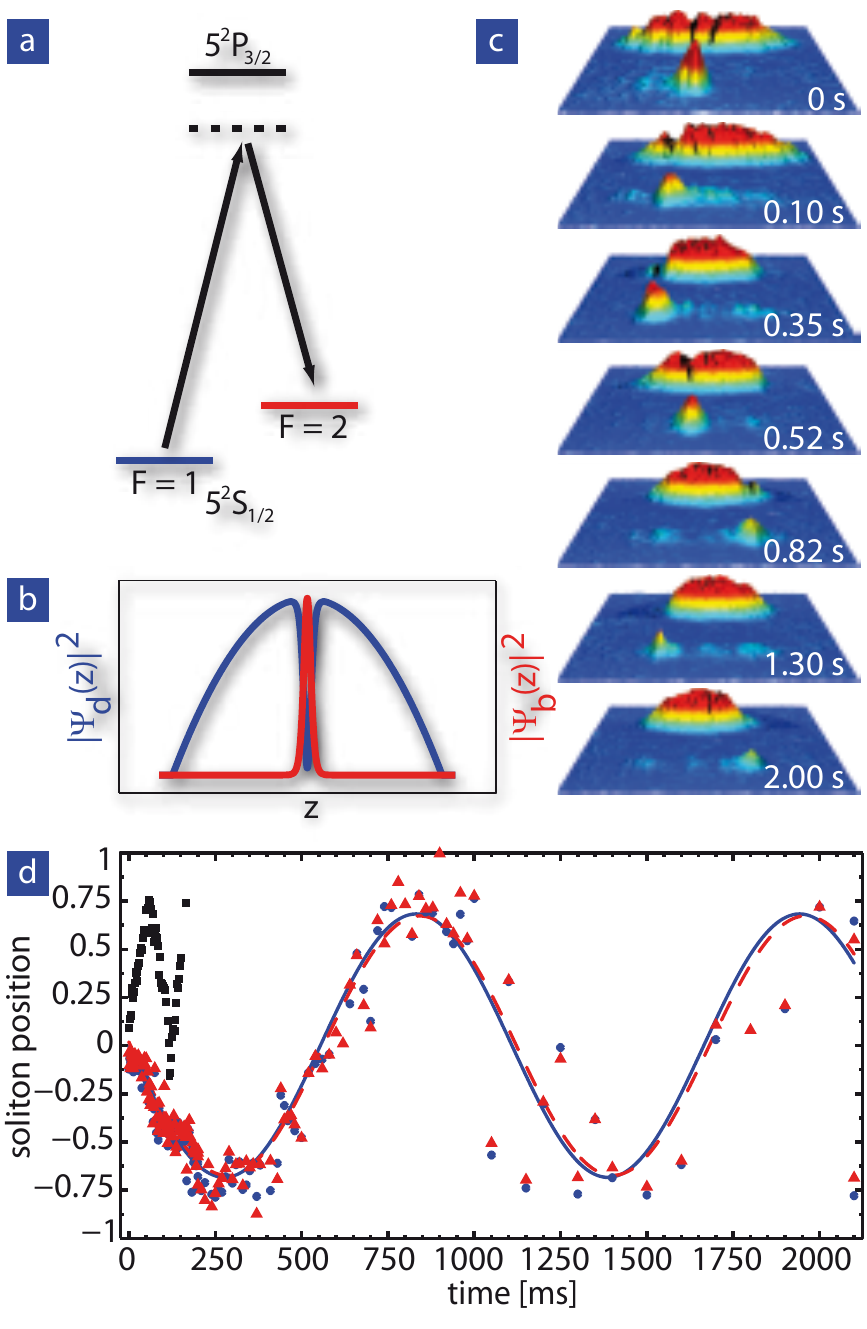}\\
  \caption[Figure3]{
	\textbf{Figure 3 | Creation and oscillation of a dark-bright soliton.}
	\textbf{a} A local population transfer in the center of the trapped BEC is achieved by a coherent two-photon 
	Raman process  between the two hyperfine states $F=1$ and $F=2$ leading to the generation of a 
	dark-bright soliton (\textbf{b}). 
	\textbf{c} A set of double exposure absorption images showing the density distributions of the two components
	which undergo slow oscillations in the axial direction.
	\textbf{d} Time series of the axial positions of the dark ({\color{blue}$\bullet$}) and bright ({\color{red}$\blacktriangle$})
	component of the soliton in addition to corresponding sinusoidal fits to the position. 
	Note that the time scale is different by almost an order of magnitude as compared to Fig.\,2c.
	For details of the first $175\,\mathrm{ms}$ see Fig.\,4.}
   \label{Figure3}
\end{figure}

\begin{figure}[f]
  \centering
  \includegraphics{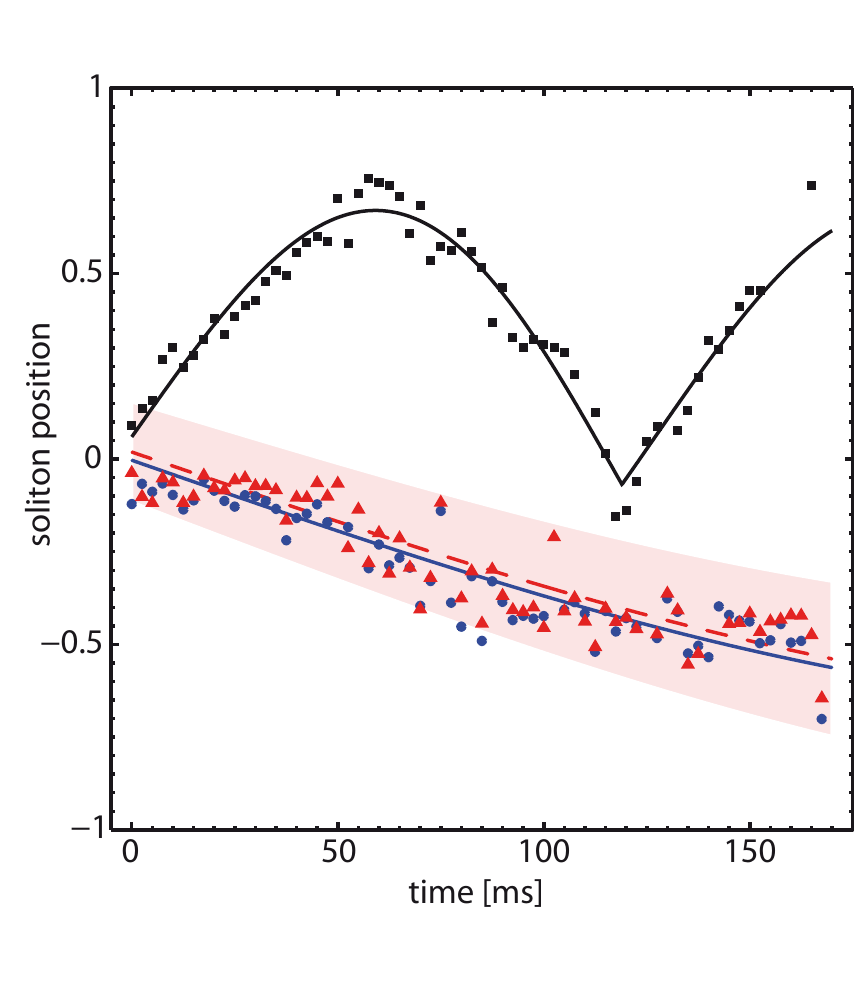}\\
  \caption[Figure4]{
\textbf{Figure 4 | Collision of a dark and a dark-bright soliton.}
	A detailed plot of the first $175\,\mathrm{ms}$ of Figure 3 reveals the reflection of an extra dark soliton off the dark-bright soliton.
	The axial position of the extra dark soliton ({\rule[0pt]{6pt}{6pt}}) is plotted together with a fit to the data (\rule[2pt]{10pt}{2pt}). 
	The mean $e^{-2}$ width of the bright soliton is indicated ({\color{lightred}\rule[0pt]{10pt}{4pt}}).
	The reflection of the extra dark soliton is very close to that expected from a hard-wall reflection. 
	The fit corresponds to a sine-function mirrored at the reflection time $t_{r}=117\,\mathrm{ms}$.}

   \label{Figure4}
\end{figure}

\end{document}